\documentclass[]{tMPH2e}
\begin{document}
\issn{1029-0435}  \issnp{0892-7022}
\jvol{112} \jnum{17} \jyear{2014} \jmonth{}

\markboth{{Stephan Werth, Gabor Rutkai, Martin Horsch, Jadran Vrabec \& Hans Hasse}}{Molecular Physics}

\articletype{}

\title{Long range correction for multi-site Lennard-Jones models and planar interfaces}

\author{Stephan Werth$^1$, Gabor Rutkai$^2$, Jadran Vrabec$^2$, Martin Horsch$^{1\ast}$\thanks{$^\ast$Corresponding author. Email: martin.horsch@mv.uni-kl.de
\vspace{6pt}} \& Hans Hasse$^1$ \\{\vbox{\vspace{12pt} {\em{$^1$Laboratory of Engineering Thermodynamics, Department of Mechanical and Process Engineering, University of Kaiserslautern,
Erwin-Schr\"odinger-Str. 44, 67663 Kaiserslautern, Germany}}
\\ {\em{$^2$Thermodynamics and Energy Technology, Department of Mechanical Engineering, University of Paderborn,
Warburger Str. 100, 33098 Paderborn, Germany}}\\\vspace{6pt}\received{submitted September 2013}}} }

\maketitle

\begin{abstract}
A slab based long range correction approach for multi-site Lennard-Jones models is presented for systems with a planar film geometry that is based on the work by Jane\v{c}ek, J. Phys. Chem. B 110: 6264 (2006). It is efficient because it relies on a center-of-mass cutoff scheme and scales in terms of numerics almost perfectly with the molecule number. For validation, a series of simulations with the two-center Lennard-Jones model fluid, carbon dioxide and cyclohexane is carried out. The results of the present approach, a site-based long range correction and simulations without any long range correction are compared with respect to the saturated liquid density and the surface tension. The present simulation results exhibit only a weak dependence on the cutoff radius, indicating a high accuracy of the implemented long range correction. 
\bigskip

\bigskip

\end{abstract}

\section{Introduction}
One of the most important properties of vapor-liquid equilibria that can be determined by molecular simulation, is the surface tension \cite{YYF10,TGWCR84,BB92,OVBBKS12}. 
Usually, the properties of interfaces are directly sampled in a simulation volume containing both the vapor and the liquid phase, separated by an interface. Indirect methods like Grand Equilibrium \cite{VH02}, $NpT$ plus test particle \cite{MF90} or Gibbs ensemble \cite{Pana02} provide access to the bulk properties along the saturation curve in a numerically more efficient manner but do not consider interfaces.

Intermolecular interactions are usually evaluated explicitly up to a specified cutoff radius, beyond which the interactions are covered by a mean field approach, i.e. a long range correction (LRC) which compensates for the cutoff \cite{SVH01,SVH03,UBDBRF00}. In homogeneous simulations, the LRC typically only considers the energy and the virial \cite{Neumann85,Onsager36}, while in inhomogeneous systems also the force has to be corrected appropriately \cite{SME07,Janecek06}. If a small cutoff radius is used without a LRC, the surface tension and other thermodynamic properties are known to deviate significantly from the correct values \cite{TA99,VKFH06}.

For homogeneous systems, typical correction strategies are straightforward, making the approximation that the 
pair correlation function is unity beyond the cutoff radius. They may rely on a site-site correction \cite{AT87} or on center-of-mass correction approaches, employing angle averaging \cite{L88} or the reaction field method \cite{SFN91,BW73}. For inhomogeneous configurations, fast multipole methods \cite{MENT03,YBKBH11}, slab based LRC \cite{LVF90,MWF97,GL97,Janecek06} or more complex Ewald summation techniques are used \cite{VIG07,IHMI12}. Recent implementations of the slab based LRC and the Ewald summation technique yield very similar results for planar interfaces \cite{Janecek06,IHMI12}.

In addition to the LRC approach, the cutoff scheme plays an important role. For molecular models consisting of several interaction sites, the computational effort is much smaller for a center-of-mass cutoff compared to a site-site cutoff. This advantage rises with the number of sites. 
However, the LRC has to be consistent with the chosen scheme \cite{L88,JKS06}. A site-site cutoff scheme consumes a much larger amount of computing time, because every site-site distance has to be evaluated and compared to the cutoff radius. E.g., for a pair of carbon dioxide models consisting of three Lennard-Jones sites each, the site-site cutoff scheme requires the execution of nine distance calculations and $if$ statements for the Lennard-Jones interactions during the neighborhood search, while a center-of-mass cutoff requires only one. Hence, a center-of-mass cutoff scheme should be preferred. 

In the present work, we combine the slab based LRC approach for inhomogeneous systems by Jane\v{c}ek \cite{Janecek06} with the center-of-mass cutoff method by Lustig \cite{L88}, which is based on angle averaging, and apply it to molecular models containing several Lennard-Jones sites. This combined correction approach is validated for planar interfaces with a two-center Lennard-Jones model fluid and two fluid models representing carbon dioxide and cyclohexane.

\section{Theory}

\begin{figure}[htb]
\centering
 \includegraphics{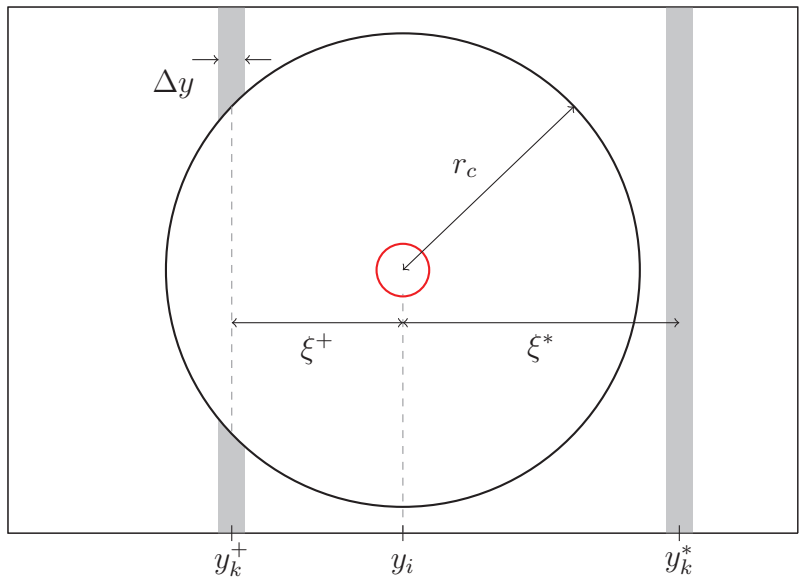}
\caption{Relevant distances for the LRC approach by Jane\v{c}ek \cite{Janecek06}. If the distance $\xi$ between the molecule $i$ and the slab $k$ is smaller than the cutoff radius, the cutoff radius has to be used as the lower integration bound, cf. Eqs. (\ref{eq:Janecek}) and (\ref{Eq:rc}).} 
\label{fig:LRC}
\end{figure}

The intermolecular pair potential $u$ is usually evaluated in molecular simulation explicitly only up to a specified cutoff radius $r_c$. To correct for the error made by this approximation, a LRC has to be applied. The potential energy of molecule $i$ is thus separated into the explicitly computed contribution and the LRC contribution
\begin{equation}
U_i = \sum_{r_{ij}< r_c} u_{ij}+ U_i^{\rm LRC}.
\end{equation}
For systems with planar symmetry, such as a planar liquid film surrounded by vapor, it is sufficient to compute the LRC in terms of the coordinate normal to the interface, employing a slab-based approach \cite{Janecek06,MWF97}. In the present work, this corresponds to the $y$ direction. The correction term $U_i^{\rm LRC}$ is then a sum over all $N_{\rm s}$ slabs with respect to the interactions $\Delta u_{i,k}^{\rm LRC} $ between the molecule $i$ and the molecules in slab $k$  
\begin{equation}
U_i^{\rm LRC} = \sum_k^{N_{\rm s}} \Delta u_{i,k}^{\rm LRC}.
\label{eq:Sum}
\end{equation}
According to Jane\v{c}ek \cite{Janecek06}, the correction term  $\Delta u_{i,k}^{\rm LRC}$ is an integral over the slab volume
\begin{equation}
\Delta u_{i,k}^{\rm LRC} = 2 \pi \rho (y_k) \Delta y \int_{r'}^\infty \text{d} r \,u(r)  r  ,
\label{eq:Janecek}
\end{equation}
where $\rho(y_k)$ denotes the mean density in slab $k$ and $\Delta y$ is the slab thickness, cf. Figure \ref{fig:LRC}. As usual, it was assumed by Jane\v{c}ek \cite{Janecek06} that for the radial distribution function within a single slab $g(r)$ $\approx$ 1 holds beyond the cutoff radius. 

According to Siperstein, the lower bound of the integration $r'$ has to be selected appropriately \cite{SMT02} as shown in Figure \ref{fig:LRC}. If the distance $\xi= | y_i - y_k |$ between the molecule $i$ and the slab $k$ is smaller than the cutoff radius, the cutoff radius has to be used as the lower integration bound, otherwise it is $\xi$ \cite{SMT02,Janecek06}, i.e.
\begin{equation}
 r' =\begin{cases}
  \xi, \quad \text{if} \quad \xi > r_c \\
  r_c, \quad \text{instead}.
 \end{cases}
 \label{Eq:rc}
\end{equation}
This definition of $r'$ has to be employed in Eq. (\ref{eq:Janecek}) as well as the analogous expressions for the force and the virial.

 Jane\v{c}ek's approach yields results that are hardly dependent on the cutoff radius for the single-site Lennard-Jones fluid down to $r_c = 2.5$ $\sigma$ \cite{Janecek06,WLHH13}. It is also suitable for multi-site models if the molecular simulation code is based on a site-site cutoff scheme. 

\begin{figure}[htb]
\centering
 \includegraphics{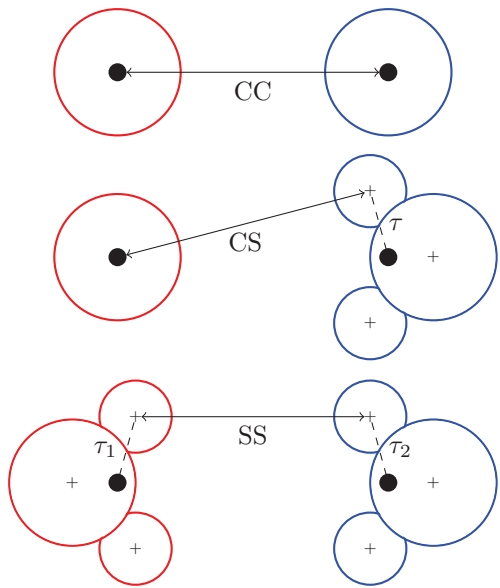}
 \caption{Illustration of the three different cases discussed here. Sites in the center of mass interact with each other as a center-center interaction (top), as opposed to the center-site interaction (middle) and the site-site interaction (bottom). The distance of the sites from the center of mass of their molecule is denoted by $\tau$. The dots indicate the center of mass, while the crosses denote the site positions.}
 \label{fig:AngleAveraging}
\end{figure}

However, for molecules consisting of several Lennard-Jones sites, a center-of-mass cutoff scheme is more efficient because only the distances between the centers of mass have to be evaluated during the neighborhood search. In this case, angle averaging as proposed by Lustig \cite{L88} is required for the LRC, because the orientation of the molecules cannot be considered explicitly by the LRC. The present study introduces such an approach, applying it to the Lennard-Jones potential
\begin{equation}
u =  4 \epsilon \left[ \sigma^{12}s^{-12}-\sigma^{6} s^{-6}\right],
\label{eq:LJ}
\end{equation}
with the energy parameter $\epsilon$ and the size parameter $\sigma$, where $s$ represents the distance between the interaction sites, which may deviate from the distance between the centers of mass $r$. Three cases have to be distinguished here, cf. Figure \ref{fig:AngleAveraging}. For a given $r$, the center-center (CC), center-site (CS) and site-site (SS) distances depend on the mutual orientation of the molecules. The term $s$ thus has to be an average over all molecular orientations with the same center-of-mass distance $r$ \cite{L88}.

\subsection*{Center-center case}
In the CC case, i.e., for the interaction between Lennard-Jones sites in the center of mass, the distance $s$ is equal to the center-of-mass distance $r$ and no angle averaging is required, because $s^{2n}=r^{2n}$. For the CC case, the reader is referred to Jane\v{c}ek \cite{Janecek06}, who derived correction terms for the potential energy, virial and force. The present work generalizes Jane\v{c}ek's approach such that a center-of-mass cutoff scheme can be applied to CS and SS interactions with a similar accuracy.

\subsection*{Center-site case}
In the CS case, a site is not in the center of mass of its molecule, i.e. it is situated at a distance $\tau$ from the center of mass. The CS case does not exist on its own, because CC and SS interactions are also always present in such a scenario. The angle-averaged value of $s^{2n}$ has been derived by Lustig \cite{L88}
\begin{equation}
s^{2n} = \frac{(r+\tau )^{2n+2}-(r-\tau )^{2n+2}}{4 r \tau (n+1)},
\label{eq:s2nCS}
\end{equation} 
where $n=-6$ or $-3$, respectively, for the repulsive or dispersive contributions to the Lennard-Jones potential. The correction term for the potential energy is then a combination of Eqs. (\ref{eq:Janecek}) and (\ref{eq:s2nCS})
\begin{align}
\Delta u_{i,k}^{\rm LRC} &= 2\pi  \rho(y_k) \Delta y \int_{r'}^\infty \text{d} r \, 4 \epsilon [ \sigma^{12} s^{-12} - \sigma^6 s^{-6 } ] r \notag  \\
&=-\frac{2 \pi \epsilon \rho(y_k) \Delta y}{\tau} \int_{r'}^\infty  \text{d} r \left[ \sigma^{12} \frac{(r+\tau )^{-10}-(r-\tau )^{-10}}{5} - \sigma^6 \frac{(r+\tau )^{-4}-(r-\tau )^{-4}}{2} \right]  \notag \\
&=-\frac{2 \pi \epsilon \rho(y_k) \Delta y \sigma^3}{3\tau}  \left[ \sigma^{9} \frac{(r'+\tau )^{-9}-(r'-\tau )^{-9}}{15} - \sigma^3 \frac{(r'+\tau )^{-3}-(r'-\tau )^{-3}}{2} \right].  
\end{align}
The correction term for the force is obtained in a similar manner
\begin{align}
\Delta f_{i,k}^{\rm LRC} &= - 2\pi  \rho(y_k) \Delta y \int_{r'}^\infty \text{d} r \frac{\partial u}{\partial r}\frac{\xi}{r} r  \notag  \\
&=-\frac{2 \pi \epsilon \rho(y_k) \Delta y \sigma^2 \xi}{\tau r'}  \left[ \sigma^{10} \frac{(r'+\tau )^{-10}-(r'-\tau )^{-10}}{5} - \sigma^4 \frac{(r'+\tau )^{-4}-(r'-\tau )^{-4}}{2} \right].  
\end{align}
The correction term for the virial is separated into its normal and tangential contribution. The normal contribution corresponds to the $y$ direction here that is perpendicular to the interface, and the tangential contribution corresponds to the $x$ and $z$ directions. The term for the virial in normal direction is analogous to the force
\begin{align}
\Delta \Pi_{N;i,k}^{\rm LRC} &= - \pi  \rho(y_k) \Delta y \int_{r'}^\infty \text{d} r \frac{\partial u}{\partial r}\frac{\xi^2}{r} r  \notag  \\
&=-\frac{\pi \epsilon \rho(y_k) \Delta y \sigma^2 \xi^2}{\tau r'}  \left[ \sigma^{10} \frac{(r'+\tau )^{-10}-(r'-\tau )^{-10}}{5} - \sigma^4 \frac{(r'+\tau )^{-4}-(r'-\tau )^{-4}}{2} \right].  
\end{align}
The term for the tangential virial is slightly more complicated
\begin{align}
\Delta \Pi_{T;i,k}^{\rm LRC} &= -\frac{1}{2}\pi  \rho(y_k) \Delta y \int_{r'}^\infty \text{d} r \frac{\partial u}{\partial r}\frac{r^2-\xi^2}{r} r  \notag  \\
&=-\frac{\pi \epsilon \rho(y_k) \Delta y \sigma^2 }{2 \tau r'}  \left[ \sigma^{10} \frac{(r'+\tau )^{-10}-(r'-\tau )^{-10}}{5} - \sigma^4 \frac{(r'+\tau )^{-4}-(r'-\tau )^{-4}}{2} \right] (r^2- \xi^2) \notag\\
& \quad -\frac{\pi \epsilon \rho(y_k) \Delta y \sigma^3 }{3 \tau}  \left[ \sigma^{9} \frac{(r'+\tau )^{-9}-(r'-\tau )^{-9}}{15} - \sigma^3 \frac{(r'+\tau )^{-3}-(r'-\tau )^{-3}}{2} \right].  
\end{align}
\subsection*{Site-site case}
In the SS case, the correction terms are of similar form. Both sites are not in the center of mass of their molecule, i.e. they are separated from it by the distances $\tau_1$ and $\tau_2$, respectively. The corresponding expression for $s^{2n}$ has also been derived by Lustig \cite{L88}
\begin{equation}
s^{2n} = \frac{(r+\tau_+)^{2n+3}-(r+\tau_-)^{2n+3}-(r-\tau_-)^{2n+3}+(r-\tau_+)^{2n+3}}{8r \tau_1 \tau_2 (n+1)(2n+3)},
\end{equation}  
with $\tau_+ = \tau_1+\tau_2$ and $\tau_- = \tau_1 - \tau_2$.
The correction terms for the potential energy, virial and force are calculated in the same way as for the CS case
\begin{align}
\Delta u_{i,k}^{\rm LRC} &=\frac{\pi \epsilon \rho(y_k) \Delta y \sigma^4}{12\tau_1 \tau_2} \bigg[ \sigma^{8} \frac{(r'+\tau_+ )^{-8}-(r'+\tau_- )^{-8}-(r'-\tau_- )^{-8}+(r'-\tau_+ )^{-8}}{30} \notag \\ & \quad - \sigma^2  \left[(r'+\tau_+ )^{-2}-(r'+\tau_- )^{-2}-(r'-\tau_- )^{-2}+(r'-\tau_+ )^{-2}\right]\bigg], 
\end{align}
\begin{align}
\Delta f_{i,k}^{\rm LRC} &= \frac{\pi \epsilon \rho(y_k) \Delta y \sigma^3 \xi}{3\tau_1 \tau_2 r'} \bigg[ \sigma^{9} \frac{(r'+\tau_+ )^{-9}-(r'+\tau_- )^{-9}-(r'-\tau_- )^{-9}+(r'-\tau_+ )^{-9}}{15} \notag \\ & \quad - \sigma^3  \frac{(r'+\tau_+ )^{-3}-(r'+\tau_- )^{-3}-(r'-\tau_- )^{-3}+(r'-\tau_+ )^{-3}}{2}\bigg], 
\end{align}
\begin{align}
\Delta \Pi_{N;i,k}^{\rm LRC} &=  \frac{\pi \epsilon \rho(y_k) \Delta y \sigma^3 \xi^2}{6\tau_1 \tau_2 r'} \bigg[ \sigma^{9} \frac{(r'+\tau_+ )^{-9}-(r'+\tau_- )^{-9}-(r'-\tau_- )^{-9}+(r'-\tau_+ )^{-9}}{15} \notag \\ & \quad - \sigma^3  \frac{(r'+\tau_+ )^{-3}-(r'+\tau_- )^{-3}-(r'-\tau_- )^{-3}+(r'-\tau_+ )^{-3}}{2}\bigg],
\end{align}
\begin{align}
\Delta \Pi_{T;i,k}^{\rm LRC} &=  \frac{\pi \epsilon \rho(y_k) \Delta y \sigma^3 }{12\tau_1 \tau_2 r'} \bigg[ \sigma^{9} \frac{(r'+\tau_+ )^{-9}-(r'+\tau_- )^{-9}-(r'-\tau_- )^{-9}+(r'-\tau_+ )^{-9}}{15} \notag \\ & \quad - \sigma^3  \frac{(r'+\tau_+ )^{-3}-(r'+\tau_- )^{-3}-(r'-\tau_- )^{-3}+(r'-\tau_+ )^{-3}}{2}\bigg] (r'^2-\xi^2) + \frac{\Delta u_{i,k}^{\rm LRC}}{2}.
\end{align}
The corrections for the normal and tangential virial are used for the pressure calculation. The surface tension $\gamma$ can be obtained from the difference between the normal and tangential contributions to the virial $\Pi_N-\Pi_T$, which is equivalent to the integral over the differential pressure $p_N-p_T$
\begin{equation}
\gamma = \frac{1}{2A} \left( \Pi_N -\Pi_T\right) = \int_{-\infty}^{\infty} \text{d} y \left( p_N - p_T \right) ,
\end{equation}
where $2A$ denotes the surface area of the two dividing surfaces \cite{Janecek06,WTRH83}.

\section{Simulations}

The above correction terms were implemented in the $ls1$ $MarDyn$ molecular dynamics code \cite{BBV11,ls12013} for an assessment of the present combination of the methods by Jane\v{c}ek \cite{Janecek06} and Lustig \cite{L88}.  The equations of motion were solved by a leapfrog integrator \cite{Fincham92} with a reduced time step of $\Delta t$ = 0.001 $\sigma \sqrt{m / \epsilon}$ for the two-center Lennard-Jones model fluid and a time step of $\Delta t = 1$ fs for the real fluids carbon dioxide and cyclohexane. Simulations were conducted in the canonical ensemble with $N=$ 16 000 molecules. The liquid phase was in the center of the simulation volume surrounded by vapor phases on both sides. The elongation of the simulation volume normal to the interface was between 60 and 80 $\sigma$ to limit the influence of finite size effects which may be significant for thin liquid films \cite{WLHH13}. A thickness of the LRC slabs of $\Delta y \approx 0.1$ $\sigma$ was used throughout. The spatial extension of the 
simulation 
volume in the other directions was at least 20 $\sigma$ to account for capillary waves  \cite{OLA05,
GJBM05,WSMB97}. For the scaling tests, the length of the simulation volume in $y$ direction was varied. The equilibration was conducted for 200 000 time steps and the production runs for 800 000 time steps. The statistical errors were estimated to be three times the standard deviation of four block averages, each over 200 000 time steps. 

The employed simulation program $ls1$ $MarDyn$ was designed for massively parallel high performance computing with systems containing a large number of molecules \cite{WHB13}. Accordingly, the implementation of the present LRC approach was designed to consume only a small amount of computing time and to scale well with the molecule number $N$ as well as with the number of processing units. The scaling behavior with respect to the number of processing units was discussed in previous publications \cite{BBV11,WHB13,HBCDFRWVH13} so that no weak or strong scaling experiments are shown here.

\section{Results}

A series of simulations for the two-center Lennard-Jones model fluid with an elongation of $L$ = $\sigma$ was carried out for different temperatures. The results are compared with those by Stoll et al. \cite{SVH032}, who employed the indirect Grand Equilibrium method where interfaces are absent, with a cutoff radius of $r_c$ = 5 $\sigma$. In addition, simulation results without any LRC are included here, representing the extreme case where long range interactions are completely neglected. To exemplify the necessity of angle averaging, Jane\v{c}ek's original approach was applied together with the center-of-mass cutoff scheme, although it was designed for a site-site cutoff scheme \cite{JKS06}. This is termed site-based approach in the following.

\begin{figure}[htb]
\centering
 \includegraphics{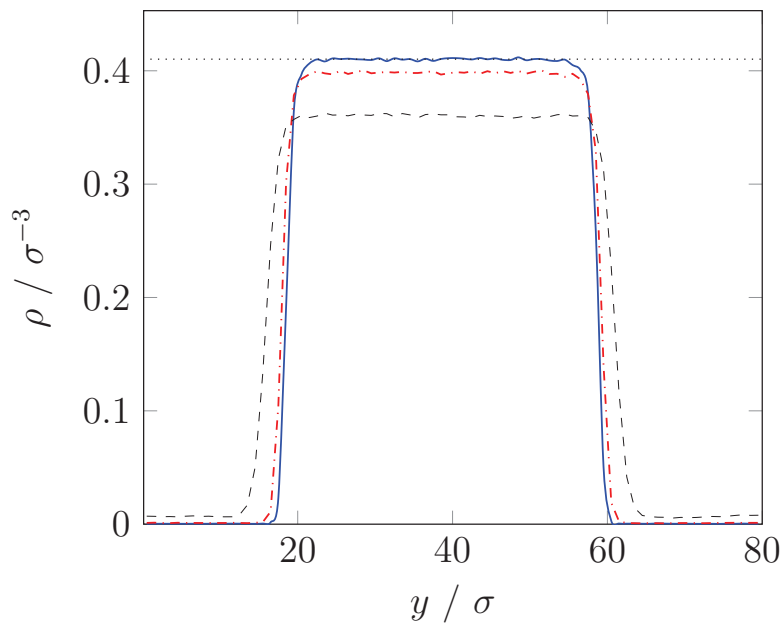}
 \caption{Density $\rho$ over $y$ coordinate for the two-center Lennard-Jones model fluid for $T$ = 0.979 $\epsilon/k_{\rm B}$ and $r_c$ = 2.5 $\sigma$. Comparison between simulations without LRC (dashed line), the site-based approach (dash-dotted line), the present approach (solid line) and the reference values by Stoll et al. \cite{SVH032} (dotted line).}
 \label{fig:2CLJ_profile}
\end{figure}

Figure \ref{fig:2CLJ_profile} shows the density over the $y$ coordinate close to the triple point. The density profile is needed on the one hand for the LRC, on the other hand for the calculation of the saturated liquid density, which is compared with results from indirect vapor-liquid equilibrium simulations. The density with the present approach matches the saturated liquid density from the homogeneous simulations by Stoll et al. \cite{SVH032}, while the other approaches exhibit deviations from the reference data.

\begin{figure}[htb]
\centering
 \includegraphics{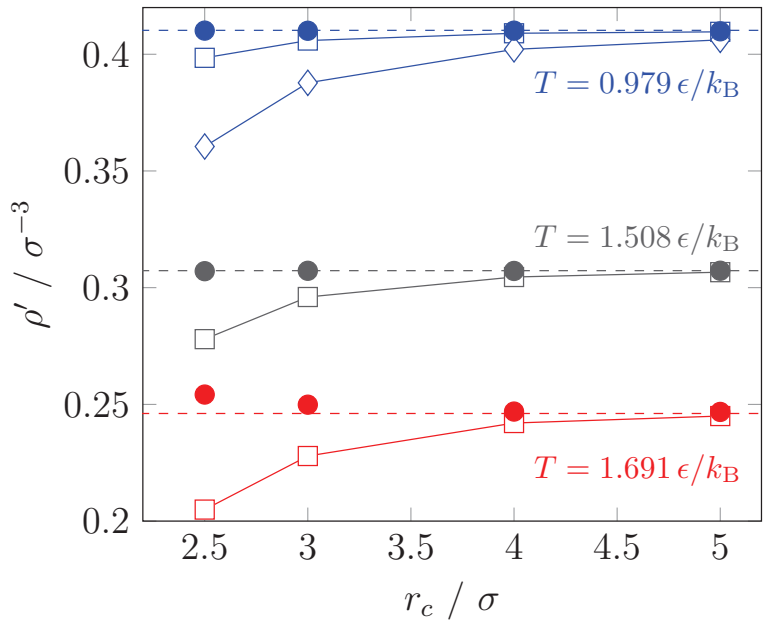}
 \caption{Saturated liquid density over the cutoff radius for the two-center Lennard-Jones model fluid. Comparison between simulations without LRC (diamonds), the site-based approach (squares), the present approach (circles) and the reference values by Stoll et al. \cite{SVH032} (dashed lines).}
 \label{fig:2CLJ_L1}
\end{figure}

Figure \ref{fig:2CLJ_L1} shows the saturated liquid density over the cutoff radius for different temperatures from near the triple point up to 0.96 $T_c$, where $T_c$ is the critical temperature. The results for the saturated liquid density that were determined with the present LRC approach hardly show any dependence on the cutoff radius for the lower two temperatures, while the site-based approach and the simulations without LRC show significant deviations from the reference saturated liquid density. Simulations without LRC were only performed for comparison near the triple point. At the highest temperature, both LRC approaches exhibit deviations from the reference case for small cutoff radii. However, it should be noted that a stable liquid film at a temperature of 0.96 $T_c$ is quite challenging to simulate due to the divergence of the correlation length at the critical point.

\begin{figure}[htb]
\centering
 \includegraphics{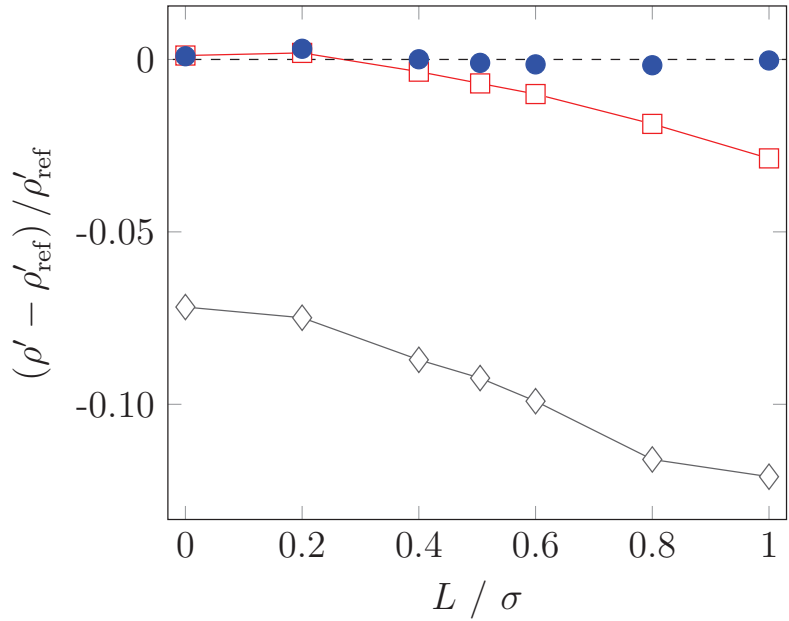}
 \caption{Relative deviations of the saturated liquid density from the reference values by Stoll et al. \cite{SVH032}. Comparison between simulations without LRC (diamonds), the site-based approach (squares) and the present approach (circles). All simulations were carried out with a cutoff radius of $r_c$ = 2.5 $\sigma$ and a temperature close to the triple point, i.e. $T$ $\approx$ $0.56$ $T_c$.}
 \label{fig:delta2CLJ}
\end{figure}

The two-center Lennard-Jones model fluid with $L$ = $\sigma$ is a difficult case, because of its anisotropy. Figure \ref{fig:delta2CLJ} shows the relative deviations from the reference data by Stoll et al. \cite{SVH032} for seven model fluids with a varied elongation $L$ at a low temperature close to their triple point. These simulations were carried out with a constant cutoff radius of $r_c$ = 2.5 $\sigma$. As expected, the deviations of the site-based approach rise with the elongation of the molecules, but they are much smaller than in case of the simulations without LRC. The deviations in terms of the saturated liquid density reach 3 \% for the site-based approach at the largest elongation $L$ = $\sigma$. For small elongations, as expected, the site-based approach and the present approach converge.

\begin{figure}[htb]
\centering
 \includegraphics{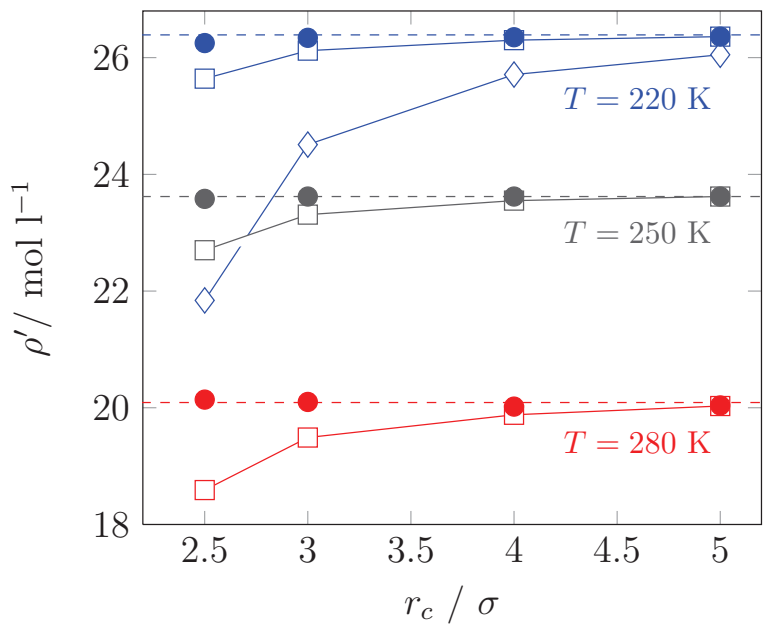}
 \caption{Saturated liquid density over the cutoff radius for carbon dioxide. Comparison between simulations without LRC (diamonds), the site-based approach (squares), the present approach (circles) and the reference values by Merker et al. \cite{MEVH10} (dashed lines).}
 \label{fig:CO2}
\end{figure}

Two multi-center Lennard-Jones fluids were also studied to compare the performance of the LRC terms: carbon dioxide (CO$_2$), which was described by a rigid three-site Lennard-Jones model with one superimposed point quadrupole \cite{MEVH10}, and cyclohexane (C$_6$H$_{12}$), which was described by a rigid six-site Lennard-Jones model \cite{MVH12}. 
For carbon dioxide, temperatures from 220 to 280 K were considered, i.e. almost from the triple point up to approximately 0.92 $T_c$. Carbon dioxide was chosen because it is similar to the two-center Lennard-Jones model fluid. Moreover, it is a combination of the CC, CS and the SS cases. 
The point quadrupole was assumed to have no preferred orientation beyond the cutoff radius, which yields a vanishing LRC contribution. 
Merker et al. \cite{MEVH10} used an indirect simulation method without the presence of interfaces and a cutoff radius of at least 7.1 $\sigma$ in terms of the Lennard-Jones parameter $\sigma$ of the oxygen atoms. Figure \ref{fig:CO2} shows the results for the saturated liquid density. The results are similar to the two-center Lennard-Jones model fluid, i.e. the saturated liquid density is almost independent on the cutoff radius with the present approach. 

\begin{figure}[htb]
\centering
 \includegraphics{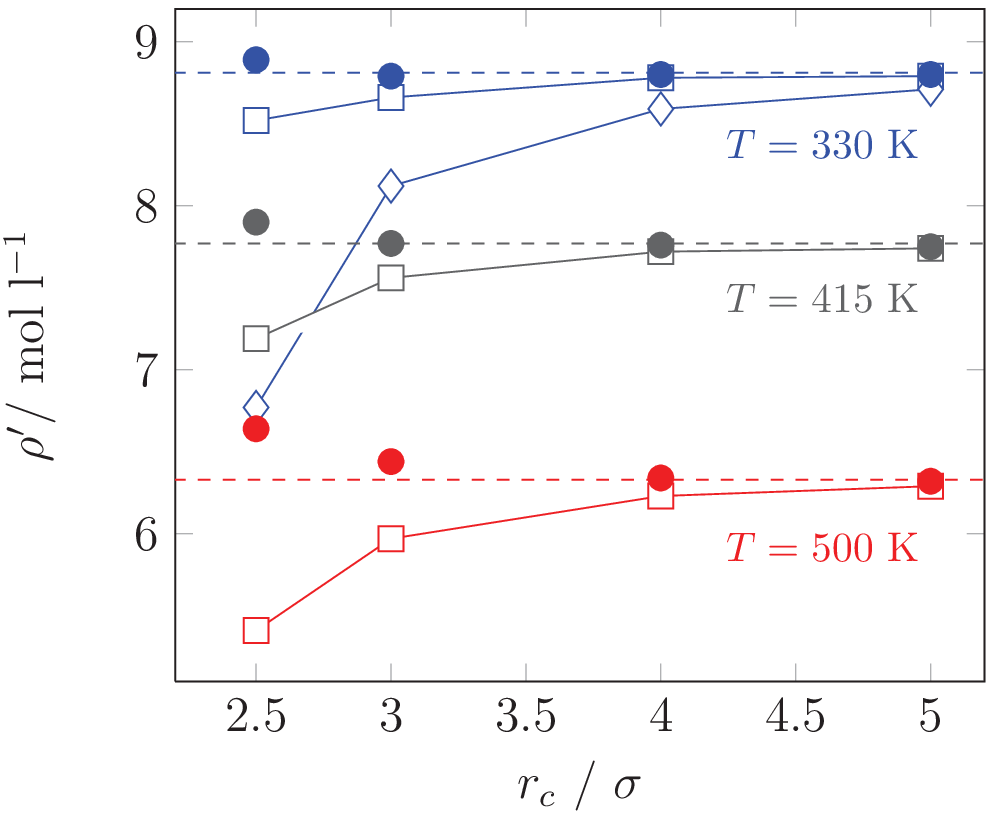}
 \caption{Saturated liquid density over the cutoff radius for the cyclohexane. Comparison between simulations without LRC (diamonds), the site-base approach (squares), the present approach (circles) and the reference values by Merker et al. \cite{MVH12} (dashed lines).}
 \label{fig:C6H12}
\end{figure}

For cyclohexane, three different temperatures were studied for a comparison between the results with different LRC approaches and  the reference data by Merker et al. \cite{MVH12}. Figure \ref{fig:C6H12} shows the results for 330, 415 and 500 K. Merker et al. \cite{MVH12} also used an indirect simulation method without the presence of interfaces and a cutoff radius of at least 4.3 $\sigma$.

Because cyclohexane is a much larger molecule than the others considered in this work, where all sites have a distance of approximately 0.52 to 0.54 $\sigma$ from the center of mass, it is obvious that a cutoff radius of 2.5 $\sigma$ is insufficient. Nonetheless, even simulations with a cutoff radius of 3 $\sigma$ yield good results in terms of the saturated liquid density. Only for a temperature of about 0.9 $T_c$, the cutoff radius must be larger.

\begin{figure}[htb]
\centering
 \includegraphics{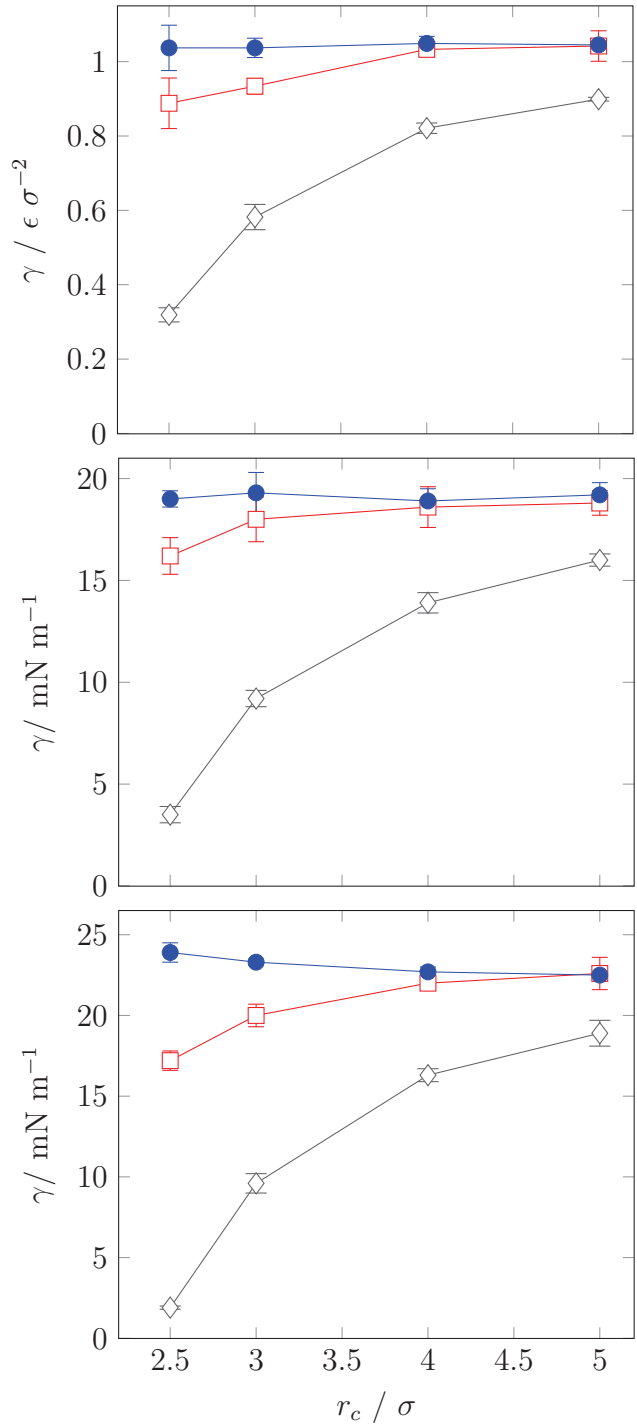}
 \caption{Surface tension over the cutoff radius for the two-center Lennard-Jones model fluid, carbon dioxide and cyclohexane. Comparison between simulations without LRC (diamonds), the site-based approach (squares) and the present approach (circles). The temperature was $T$ = 0.979 $\epsilon/k_{\rm B}$ for the two-center Lennard-Jones model fluid (top), 220 K for carbon dioxide (center) and 330 K for cyclohexane (bottom).}
 \label{fig:Gamma}
\end{figure}

Another important property of vapor-liquid equilibria is the surface tension. For the lowest temperature of the fluids discussed above, the surface tension was determined with the three different approaches. Figure \ref{fig:Gamma} shows the surface tension over the cutoff radius for the two-center Lennard-Jones model fluid, carbon dioxide and cylcohexane. The number of time steps was enlarged to four million to reduce the statistical uncertainties and better identify systematic deviations. 

The dependence of the surface tension on the cutoff radius is similar to the dependence of the density on the cutoff radius. The present approach shows hardly any influence of $r_c$ on the surface tension, as opposed to the other discussed approaches, which exhibit a significant cutoff radius dependence.

\begin{figure}[htb]
\centering
 \includegraphics{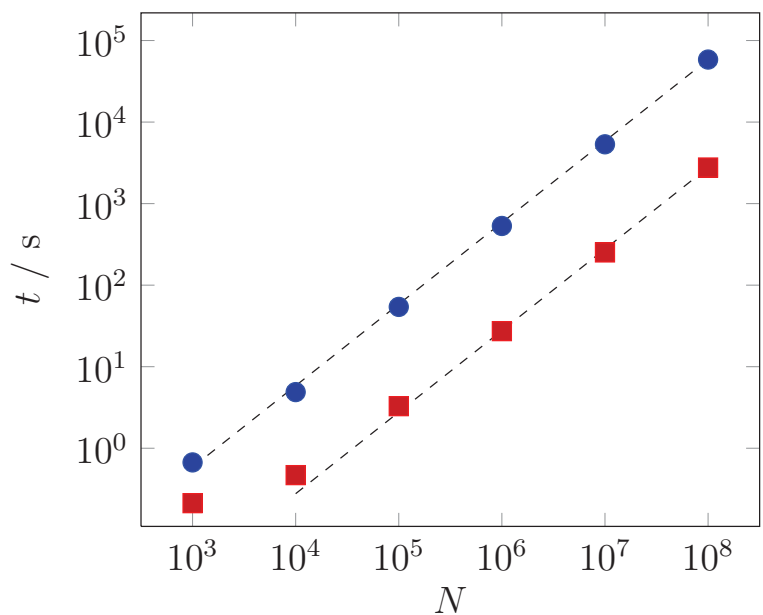}
 \caption{Computing time for 100 time steps with a single processing unit over the molecule number $N$. The circles correspond to the computing time for the explicitly evaluated interactions. The computing time for the LRC (squares) is considerably smaller.}
 \label{fig:Scaling}
\end{figure}

Furthermore, a simulation series with a single processing unit (Intel Xeon E5-2670) with a varying number of two-center Lennard-Jones molecules with an elongation $L$ = $\sigma$ was carried out. The chosen temperature $T$ = $0.979$ $\epsilon$/$k_{\rm B}$ is close to the triple point of this fluid \cite{SVH032}. Figure \ref{fig:Scaling} shows the computing time for 100 time steps in the canonical ensemble. Due to the underlying linked-cell algorithm \cite{QB73,HE81}, the computing time for the explicitly evaluated interactions scales almost perfectly with the molecule number $N$. Only for small systems below $N\approx 10^4$, the LRC does not perfectly scale with the molecule number, because the number of slabs does not correlate with it, but rather with the length of the simulation volume in $y$ direction. However, even in this case, the computational effort for the LRC is more than one order of magnitude smaller than for the explicitly evaluated interactions.

\section{Conclusion}

In this work, a new slab based LRC approach for inhomogeneous systems with planar interfaces was presented. It was applied to molecular models consisting of several Lennard-Jones interaction sites, employing a center-of-mass cutoff. The center-of-mass cutoff scheme is numerically more efficient than a site-site cutoff scheme, but it requires a more demanding LRC. The LRC by Jane\v{c}ek \cite{Janecek06} that is based on the site-site cutoff scheme was generalized to the center-of-mass cutoff scheme with the angle averaging method by Lustig \cite{L88}. The influence of the LRC on the saturated liquid density and the surface tension was studied. The present LRC approach yields very good results for both properties and shows only a weak dependence on the cutoff radius. It is numerically efficient, consumes only a small amount of computing time and scales well for systems with very large numbers of molecules.

\section*{Acknowledgment}

The authors gratefully acknowledge financial support from Deutsche Forschungsgemeinschaft (DFG) within the Collaborative Research Center (SFB) 926 as well as the project VR6/9-1 ``Thermodynamik von Tropfen unter extremen Bedinungen mittels molekularer Simulation''. They thank Thorsten Merker for providing additional information and Thomas Werth for fruitful discussions. The present work was conducted under the auspices of the Boltzmann-Zuse Society of Computational Molecular Engineering (BZS) and the simulations were carried out on the Regional University Computing Center Kaiserslautern (RHRK) under the grant TUKL-MSWS.

\label{lastpage}


\end{document}